\documentclass{article}
\pdfoutput=1
\usepackage{lscape}
\usepackage{lipsum}
\usepackage{enumerate}
\usepackage{amsfonts}
\usepackage{amsmath}
\usepackage{comment}
\usepackage{textcomp}
\usepackage{amssymb}
\usepackage{color}
\usepackage{graphicx}
\usepackage{pdflscape}
\usepackage{afterpage}
\usepackage[matrix,arrow,curve]{xy}
\usepackage{array}

\def\be{\begin{eqnarray}}
\def\ee{\end{eqnarray}}
\def\nn{\nonumber}

\makeatletter
\renewcommand*{\@cite@ofmt}{\bfseries\hbox}
\makeatother

\def\l[{\phantom.[}

\hoffset= - 1.2in         % switch off for draft style
\begin{document}

\title{\vspace{.1cm}{\Large {\bf {On $6j$-symbols for symmetric representations of $U_q(\mathfrak{su}_N)$}}\vspace{.2cm}}}
\author{
{\bf A.Mironov$^{a,b,c,d}$}\footnote{mironov@lpi.ru; mironov@itep.ru},
\ {\bf A.Morozov$^{b,c,d}$}\thanks{morozov@itep.ru},
\ \ and
 \ {\bf A.Sleptsov$^{b,c,d,e}$}\thanks{sleptsov@itep.ru}}
\date{ }
%}

\maketitle

\vspace{-5cm}

\begin{center}
\hfill FIAN/TD-18/17\\
\hfill IITP/TH-14/17\\
\hfill ITEP/TH-23/17
\end{center}

\vspace{2.3cm}

\begin{center}
$^a$ {\small {\it Lebedev Physics Institute, Moscow 119991, Russia}}\\
$^b$ {\small {\it ITEP, Moscow 117218, Russia}}\\
$^c$ {\small {\it Institute for Information Transmission Problems, Moscow 127994, Russia}}\\
$^d$ {\small {\it National Research Nuclear University MEPhI, Moscow 115409, Russia }}\\
$^e$ {\small {\it Laboratory of Quantum Topology, Chelyabinsk State University, Chelyabinsk 454001, Russia }}

\end{center}

\vspace{.5cm}

\begin{abstract}
Explicit expressions are found for the $6j$ symbols in symmetric
representations of quantum $\mathfrak{su}_N$
through appropriate hypergeometric Askey-Wilson (q-Racah) polynomials.
This generalizes the well-known classical formulas for $U_q(\mathfrak{su}_2)$
and provides a link to conformal theories and matrix models.
\end{abstract}

\bigskip

\section{Introduction}

The theory of Racah-Wigner coefficients (6j-symbols) \cite{RW} is among the standard
topics in theoretical physics textbooks including the celebrated
{\it Quantum Mechanics} of L.Landau and E.Lifshitz \cite{LL3}.
It is of course a well known story in representation theory,
because the $6j$-symbols intertwine the triple tensor products of representations,
$$
(R_1\otimes R_2)\otimes R_3  \longrightarrow R_4
$$
and
$$
R_1\otimes (R_2\otimes R_3) \longrightarrow R_4
$$
They are matrices
$\left\{\begin{array}{ccc} R_1 & R_2 & R_i \\ R_3 & R_4 & R_j \end{array}\right\}$,
with $i$ and $j$ labeling representations in the channels $R_1\otimes R_2 =\oplus_i R_i$
and $R_2\otimes R_3 =\oplus_j R_j$ respectively:

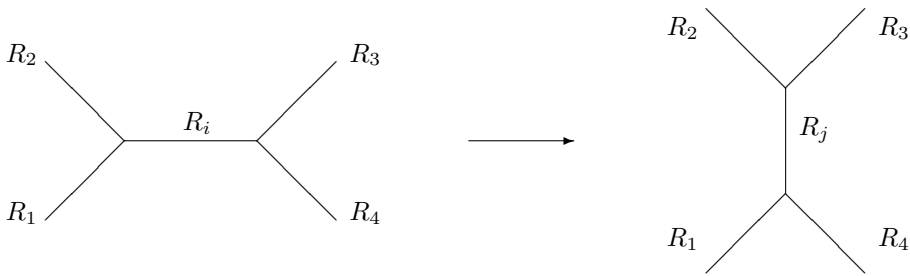
\begin{picture}(250,120)(-100,-60)
\put(0,0){\line(1,0){50}}
\put(0,0){\line(-1,1){30}}
\put(0,0){\line(-1,-1){30}}
\put(50,0){\line(1,1){30}}
\put(50,0){\line(1,-1){30}}
\put(-45,-30){\mbox{$R_1$}}
\put(-45,30){\mbox{$R_2$}}
\put(85,-30){\mbox{$R_4$}}
\put(85,30){\mbox{$R_3$}}
\put(22,4){\mbox{$R_i$}}
\put(130,0){\vector(1,0){40}}
\put(250,0){
\put(0,-20){\line(0,1){40}}
\put(0,-20){\line(-1,-1){30}}
\put(0,-20){\line(1,-1){30}}
\put(0,20){\line(1,1){30}}
\put(0,20){\line(-1,1){30}}
\put(-45,-40){\mbox{$R_1$}}
\put(-45,40){\mbox{$R_2$}}
\put(35,40){\mbox{$R_3$}}
\put(35,-40){\mbox{$R_4$}}
\put(5,2){\mbox{$R_j$}}
}
\end{picture}

\noindent
Tensor products are widely used in different topics of theoretical
and mathematical physics from quantum mechanics to knot theory.
Often needed are explicit formulas,
because one typically wants to explicitly construct either the particle states
or solutions to Yang-Baxter equations, i.e. the quantum ${\cal R}$-matrices.
Therefore the Racah matrices were a subject of intensive investigation
during the last three decades \cite{Rbook}, still surprisingly few results were
obtained, until the very recent advances \cite{Racah,GJ,MMSpret,MMMS,Mor}\footnote{Some of the quantum Racah matrices for $\mathfrak{su}_N$ are available at \cite{knotebook}.}, which came from the newly discovered
arborescent calculus \cite{Kaul,NRZ,GMMMS,MMSpret,arbor,MMMS}
and differential expansions \cite{IMMMfe,diffexpan,Sulk} of knot polynomials.
These approaches allowed one to calculate many Racah matrices in various representations,
but they are not yet brought into analytic form,
i.e. all matrix elements are  explicitly listed, but not described by a general
formula with arbitrary $i$ and $j$.
In fact, getting such analytic formulas appears to be a separate non-trivial problem,
and, in this letter, we address it in the very simple case of symmetric representations
$R_i$,
described by the single-line Young diagrams $[r_i]$ of length $r_i$,
and their $\mathfrak{su}_N$-conjugates $\bar R_i$, described by the diagrams $[r_i^{N-1}]$
with $N-1$ lines of the same length.
Somewhat surprisingly even in this case the answer was long known for $\mathfrak{su}_2$,
but not for generic $\mathfrak{su}_N$.
We perform this extension from $2$ to $N$ and use this example to describe the main
ideas, which can hopefully lead to generalizations for non-symmetric representations
(pure antisymmetric case is related to pure symmetric by the simple transformation $q\to -1/q$).

The key point is \cite{AW,KK,Klimyk-Vilenkin,Kirilov-Reshetikhin}
that the quantum 6j-symbols for $U_q(\mathfrak{su}_2)$
can be expressed through the balanced $q$-hypergeometric series $_4\phi_3$.
There are many ways to do this, but one is distinguished,
because it involves the q-orthogonal polynomials
(often named Racah polynomials, which are a particular case of the Askey-Wilson polynomials).
We explain how these formulas are deformed from $N=2$ to arbitrary $N\geq 2$,
and discuss their properties, in particular the 3-term relations,
which are the necessary property of orthogonal polynomials,
and their connection to group theory pentagon identities \cite{penta}
and to the Koornwinder-Macdonald polynomials \cite{Mac,Koor}.

\bigskip

\section{Quantum 6-j symbols for symmetric representations}

In \cite{Racah,MMSpret}, there  were obtained analytic formulas
for the two kinds of quantum 6j-symbols for arbitrary symmetric representations
\be
\text{(I  kind)} \
\bar S = \epsilon_{\{R_i\}}\sqrt{\text{dim}_q\,R_{12} \ \text{dim}_q\,R_{23}} \cdot \left\{
\begin{array}{ccc}
R_1&\bar R_2&R_{12} \\
R_3&\bar R_4&R_{23}
\end{array}
\right\}, \nonumber \\ \text{(II kind)} \
S = \epsilon_{\{R_i\}}\sqrt{\text{dim}_q\,R_{12} \ \text{dim}_q\,R_{23}} \cdot \left\{
\begin{array}{ccc}
R_1& R_2&R_{12} \\
\bar R_3&\bar R_4&R_{23}
\end{array}
\right\},
\ee
where $\epsilon_{\{R_i\}} = \pm 1$ and everywhere below it is given by $(-1)^{i+j+r}$, while $\text{dim}_q\,R$ is a quantum dimension of representation $R$.
These two matrices naturally arise in arborescent knot calculus.
They are unitary and are related via
\be
\bar S = \bar T^{-1} S T^{-1} S^\dagger \bar T^{-1}
\label{SvsbS}
\ee
with the diagonal matrices $T$ and $\bar T$ made from the eigenvalues of relevant ${\cal R}$-matrices,
\be
T = {\rm diag}\left((-1)^{m+1}\dfrac{q^{-r^2+m^2+m}}{A^r}\right), \ m=0..r \nn\\
\bar T = {\rm diag}\Big((-q^{m-1}A)^m\Big), \ m=0..r
\ee

\paragraph{Racah matrices of kind I.} We begin with the matrices of the first kind,
when $R_1=R_3=[r]$, $R_2=R_4=[\bar r]$ so that $R_{12}$ and $R_{23}$ are representations of the type $\mathfrak{R}_n=[2n,n^{N-2}]$ of $\mathfrak{su}_N$, which emerge in the decomposition of $[r]\otimes[\bar r]$:
\be
[r]\otimes[\bar r]=\oplus_{n=0}^r\mathfrak{R}_n
\ee
We denote $R_{12}=\mathfrak{R}_i\to i$ and $R_{23}=\mathfrak{R}_j\to j$ and the Racah matrix is symmetric in $i$ and $j$.
In this case, the formulas were given in \cite{MMSpret} and \cite{Racah}:
\be
\label{f1}
\left\{
\begin{array}{ccc}
r&\bar r&i \\
r&\bar r&j
\end{array}
\right\} = \dfrac{[i]!^2[j]!^2[r{-}i]![r{-}j]![N{-}1]![N{-}2]!}{[r+i+N-1]![r+j+N-1]!} \sum_z (-)^z \dfrac{[r+N-1+z]!}{[z{-}i]!^2[z{-}j]!^2[r{-}z]![i{+}j{-}z]![i{+}j{+}N{-}2{-}z]!},
\ee
where $[n]:=\dfrac{q^n-q^{-n}}{q-q^{-1}}$ is a quantum number. Making the transformation $z=r-s$, one gets
\be
\label{f2}
\left\{
\begin{array}{ccc}
r&\bar r&i \\
r&\bar r&j
\end{array}
\right\} = \dfrac{[i]!^2[j]!^2[r{-}i]![r{-}j]![N{-}1]![N{-}2]!}{[r+i+N-1]![r+j+N-1]!} \sum_s (-)^{r-s} \dfrac{[2r+N-1-s]!}{[r{-}s{-}i]!^2[r{-}s{-}j]!^2[s]![i{+}j{-}r+s]![i{+}j{+}N{-}2{-}r{+}s]!}
\ee

Introducing the q-hypergeometric series
\be
_{p+1}\phi_p\left[
\begin{array}{c|c}
a_1, \ldots, a_{p+1} & \\
& q,z \\
b_1, \ldots, b_p &
\end{array}
\right] = \sum_{n=0}^{\infty} \dfrac{(a_1,\ldots ,a_{p+1};q)_n}{(b_1,\ldots ,b_p;q)_n (q,q)_n}z^n, \\
(a;q)_n = \prod_{k=0}^{n-1}(1-aq^k),\ \ \ \ \ \ \ (a_1,\ldots ,a_i;q)_n=(a_1;q)_n\ldots (a_i;q)_n \\
_{p+1}\Phi_p\left[
\begin{array}{c|c}
a_1, \ldots, a_{p+1} & \\
& q,z \\
b_1, \ldots, b_p &
\end{array}
\right] = {}_{p+1}\phi_p\left[
\begin{array}{c|c}
q^{a_1}, \ldots, q^{a_{p+1}} & \\
& q,z \\
q^{b_1}, \ldots, q^{b_p} &
\end{array}
\right]
\ee
one can write formula (\ref{f2}) as
\be
\label{f3}
\left\{
\begin{array}{ccc}
r&\bar r&i \\
r&\bar r&j
\end{array}
\right\}
= \dfrac{[i]!^2[j]!^2[2r{+}N{-}1]![N{-}1]![N{-}2]!}
{[r{+}i{+}N{-}1]![r{+}j{+}N{-}1]![r{-}i]![r{-}j]![i{+}j{-}r]![i{+}j{-}r{+}N{-}2]!}   \cdot
\nn \\
\cdot {_4\Phi_3}
\left[
\begin{array}{c|c}
i{-}r,i{-}r,j{-}r,j{-}r & \\
& q^2, q^2\\
1{-}2r{-}N,i{+}j{-}r{+}1,i{+}j{-}r{+}N{-}1 &
\end{array}
\right]
\ee

In order to obtain alternative expressions for the 6j-symbols, one can use Sears' transformations for terminating balanced $_4\Phi_3\left[\ldots;q^2,q^2\right]$ (see \cite{GR,Klimyk-Vilenkin}):
\be
\label{transf}
{_4\Phi_3}
{\footnotesize
\left[
\begin{array}{c|c}
x,y,z,n & \\
& q^2, q^2\\
u,v,w &
\end{array}
\right]
}    =  \dfrac{[v{-}z{-}n{-}1]![u{-}z{-}n{-}1]![v{-}1]![u{-}1]!}{[v{-}z{-}1]![v{-}n{-}1]![u{-}z{-}1]![u{-}n{-}1]!} \cdot {_4\Phi_3}
\footnotesize{
\left[
\begin{array}{c|c}
w-x,w-y,z,n & \\
& q^2, q^2\\
1-u+z+n,1-v+z+n,w &
\end{array}
\right]
}
\ee
where the numbers $x,y,z,u,v,w$ are integral and the balanced series condition
\be
x+y+z+n+1 = u+v+w
\label{balance}
\ee
is fulfilled. Also one can use the invariance of $_4\Phi_3\left[\ldots;q^2,q^2\right]$ under permutations of $x,y,z,n$ or $u,v,w$. Then, one gets the following expression for the $6j$-symbols:
\be
\!\!\!\!\!\!\!
\left\{
\begin{array}{ccc}
r&\bar r&i \\
r&\bar r&j
\end{array}
\right\}
=
\dfrac{[i]![j]![N{-}1]![N{-}2]!}{[i{+}N{-}2]![j{+}N{-}2]!} \dfrac{[r]![r+N-2]![2r+N-1]!}{[r{-}i]![r{-}j]![r{+}i{+}N{-}1]![r{+}j{+}N{-}1]!}
\cdot   {_4\Phi_3}
{\tiny
\left[
\begin{array}{c|c}
j{-}r,{-}r{-}j{-}N{+}1,i{-}r,{-}r{-}i{-}N{+}1 & \\
& q^2, q^2\\
-r,{-}r{-}N{+}2,{-}2r{-}N{+}1 &
\end{array}
\right]
}
\label{hg3}
\ee
or expanding $_4\Phi_3\left[\ldots;q^2,q^2\right]$ in terms of q-factorials, one gets a generalization of the most commonly encountered formulas in the literature about $\mathfrak{su}_2$ 6j-symbols:
\be
\label{f1}
\left\{
\begin{array}{ccc}
r&\bar r&i \\
r&\bar r&j
\end{array}
\right\} &=& \dfrac{[i]!^2[j]!^2[r{-}i]![r{-}j]![N{-}1]![N{-}2]!}{[r+i+N-1]![r+j+N-1]!} \sum_z (-)^z \dfrac{[r+N-1+z]!}{[z{-}i]!^2[z{-}j]!^2[r{-}z]![i{+}j{-}z]![i{+}j{+}N{-}2{-}z]!}  \\
&=& \dfrac{[i]![j]![N-1]![N-2]!}{[i+N-2]![j+N-2]!} \sum_z (-)^z \dfrac{[r+N-2-z]![z+i]![z+j]!}{[z]![r-i-z]![r-j-z]![i{+}j{-}r{+}z]![i{+}j{+}N{-}1{+}z]!} \\
&=& \dfrac{[i]![j]![N-1]![N-2]!}{[i+N-2]![j+N-2]!} \sum_z (-)^z \dfrac{[r+N-2-z]![r-z]![2r+N-1-z]!}{[z]![r-i-z]![r-j-z]![r{+}i{+}N{-}1{-}z]![r{+}j{+}N{-}1{-}z]!}
\label{f3}
\ee
These three expansions are related by Sears' transformations (\ref{transf})
of the balanced hypergeometric
series.

\bigskip

Now let us define the q-Racah polynomial in variable $\mu(x)=q^{-x}+q^{x+c+d+1}$ of degree $n$ as \cite{GR}
\be
{\cal R}_n\left(\mu(x)\,\Big|\, a,b,c,d\,\Big|\, q\right) := {}_4\Phi_3\left[
\begin{array}{c|c}
-n, a+b+n+1, -x, x+c+d+1 & \\
& q,q \\
a+1, b+d+1, c+1 &
\end{array}
\right],
\label{Racahpol}
\ee
where $a+1=-m$ or $b+d+1=-m$ or $c+1=-m$, $m\in\mathbb{Z}_+$. Since
\be
(q^{-x};q)_j\cdot(q^{x+c+d+1};q)_j = \prod_{k=0}^{j-1}(1-q^{-x+k})(1-q^{x+c+d+1+k}) = \prod_{k=0}^{j-1} (1 - \mu(x)q^k + q^{c+d+1+2k}),
\ee
it is clear that $R_n\left(\mu(x); a,b,c,d; q\right)$ is a polynomial of degree $n$ in $\mu(x)$.
In terms of the $q$-Racah polynomials (\ref{Racahpol}), formula (\ref{hg3}) takes the form
{\footnotesize
\be
\boxed{
\left\{
\begin{array}{ccc}
r&\bar r&i \\
r&\bar r&j
\end{array}
\right\}
\ \ = \ \ \dfrac{[i]![j]![N{-}1]![N{-}2]!}{[i{+}N{-}2]![j{+}N{-}2]!} \dfrac{[r]![r+N-2]![2r+N-1]!}{[r{-}i]![r{-}j]![r{+}i{+}N{-}1]![r{+}j{+}N{-}1]!} \cdot
  {\cal R}_{r-j}\left(q^{2(i-r)}{+}q^{2(1{-}r{-}i{-}N)}\,\Big|\, {-}r{-}1, 1{-}r{-}N, {-}2r{-}N, 0\,\Big|\, q^2\right)
}
\label{qrac}
\ee
}

\paragraph{Racah matrices of kind II.} Now let us consider the matrices of the second kind, when $R_1=R_2=[r]$, $R_3=R_4=[\bar r]$
so that $R_{23}$ is still of the type $\mathfrak{R}_n=[2n,n^{N-2}]$ since belongs to the decomposition of $[r]\otimes [\bar r]$, while $R_{12}$ belongs to the decomposition
\be
[r]\otimes[r]=\oplus_{n=0}^r[r+n,r-n]
\ee
We denote $R_{12}=[r+i,r-i]\to i$ and $R_{23}=\mathfrak{R}_j\to j$ and the Racah matrix is no longer symmetric in $i$ and $j$.
Below we assume that $i<j$, otherwise one needs to change $i\longleftrightarrow j$ in all answers. In this case, the manifest formulas also can be found in \cite{MMSpret} and \cite{Racah}:
\be
\label{ff1}
\left\{
\begin{array}{ccc}
r& r&i \\
\bar r&\bar r&j
\end{array}
\right\} = \dfrac{[i]!^2[j]!^2[r{-}i]![r{-}j]![N{-}1]![N{-}2]!}{[r+i+N-1]![r+j+N-1]!} \sum_z (-)^z \dfrac{[r+N-1+z]!}{[z{-}i]!^2[z{-}j]![z-j+N-2]![r{-}z]![i{+}j{-}z]!^2} = \\
= \dfrac{[i]!^2[j]!^2[N{-}1]![N{-}2]![2r{+}N{-}1]}{[r{+}i{+}N{-}1]![r{+}j{+}N{-}1]! [r{-}i]![r{-}j{+}N{-}2]![i{+}j{-}r]!^2} \cdot {_4}\Phi_3\left[
\begin{array}{c|c}
j{-}r,j{-}r{-}N{+}2,i{-}r,i{-}r & \\
& q^2, q^2\\
i{+}j{-}r{+}1,i{+}j{-}r{+}1,1{-}N{-}2r &
\end{array}
\right]
\ee
With the help of Sears' transformation (\ref{transf}), one can convert this into
\be
\label{hg4}
\left\{
\begin{array}{ccc}
r& r&i \\
\bar r&\bar r&j
\end{array}
\right\} =
\dfrac{[r]!^2[N{-}1]![N{-}2]![2r+N-1]!}{[r{-}i]![r{-}j{+}N{-}2]![r{+}i{+}N{-}1]![r{+}j{+}N{-}1]!} \cdot {_4\Phi_3}
\left[
\begin{array}{c|c}
j{-}r,{-}r{-}j{-}1,i{-}r,{-}r{-}i{-}N{+}1 & \\
& q^2, q^2\\
-r,{-}r,{-}2r{-}N{+}1 &
\end{array}
\right]
\ee
and, in terms of the q-Racah polynomials (\ref{Racahpol}), one has
{\footnotesize
\be
\label{qr4}
\!\!\!\!\!\!\!\!\!\!\!\!\!\!\!\!
\boxed{
\left\{
\begin{array}{ccc}
r& r&i \\
\bar r&\bar r&j
\end{array}
\right\} =
\dfrac{[r]!^2[N{-}1]![N{-}2]![2r+N-1]!}{[r{-}i]![r{-}j{+}N{-}2]![r{+}i{+}N{-}1]![r{+}j{+}N{-}1]!} \cdot {\cal R}_{r-j}\left(q^{2(i-r)}{+}q^{2(1{-}r{-}i{-}N)}\,\Big|\, {-}r{-}1, {-}r{-}1, {-}2r{-}N, 0\,\Big|\, q^2\right)
}
\ee}

\section{Relation to Askey-Wilson and Racah polynomials}

Thus, (\ref{qrac}) and (\ref{qr4}) form the two 2-parametric ($N$, $r$) sub-varieties
\be\label{R2}
{\cal R}_{r-j}\left(q^{2(i-r)}{+}q^{2(1{-}r{-}i{-}N)}\,\Big|\, {-}r{-}1, {-}r{-}1-(N-2)p, {-}2r{-}N, 0\,\Big|\, q^2\right)
\ee
with $p=0,1$
in the $6$-dimensional set of the $q$-hypergeometric polynomials
made from the 7-parametric function $\phantom._4\phi_3(z)$,
with one of the parameters being a negative integer
(to make the hypergeometric series a polynomial)
converted into polynomial's degree
($q$ is not counted as a parameter in these terms). These polynomials are orthogonal, and they can be embedded into a large family of orthogonal polynomials.

For a system of $q$-hypergeometric polynomials to be orthogonal,
they should satisfy a
3-term relation, what requires some art and imposes additional restrictions. In the case of $\phantom._4\phi_3(z)$, it is fixing $z$ and
the balanced series condition (\ref{balance}).
In result, one gets a generic 4-parametric family of
the Askey-Wilson $q$-polynomials defined as
\be\label{AV}
P_n(X)=(ab,ac,ad;q)_na^{-n}\cdot \!\!\!\!\phantom{A}_4\phi_3\left[
\begin{array}{c|c}
q^{-n},abcdq^{n-1},ae^{i\theta},ae^{-i\theta}&\\
& q, q\\
ab,ac,ad &
\end{array}
\right]
\ee
Note that the variable $z$ of the q-hypergeometric function $\phantom._4\phi_3(z)$ is fixed at $z=q$ and $P_n(X)$ are polynomials in $X=\cos\theta$.
The standard Racah polynomials (\ref{Racahpol}) form a sub-family within this family, they can selected out by an integrality condition
$bdq=q^{-m}$, $m\in\mathbb{Z}_+$.
It slightly differs from (\ref{AV}) in normalization and
specification of variables: ${\cal R}_n\left(\mu(x)\,\Big|\, \tilde a,\tilde b,\tilde c,\tilde d\,\Big|\, q\right) $ in (\ref{Racahpol}) is a polynomials in $\tilde X=\mu(x)=q^{-x}+q^{x+\tilde c+\tilde d+1}=2aX$ with the parameters $q^{\tilde a}=ab/q$, $q^{\tilde b}=cd/q$, $q^{\tilde c}=ac/q$, $q^{\tilde d}=q/c$. In particular, (\ref{R2}) is a polynomial in $q^{2(i-r)}+q^{2(1-r-i-N)}$.

\section{Three-term relations}

The q-Racah polynomials (\ref{Racahpol}) are orthogonal with respect to the discrete measure (Jackson integral):
\be
\sum_{x=0}^{m} \dfrac{(q^{a+1},q^{b+d+1},q^{c+1},q^{c+d+1};q)_x}{(q,q^{c+d+1-a},q^{c+1-b},q^{d+1};q)_x}\dfrac{1-q^{c+d+2x+1}}{q^{x+ax+bx}(1-q^{c+d+1})}\cdot {\cal R}_n\left(\mu(x)\right) {\cal R}_k\left(\mu(x)\right) = h_n\delta_{n,k},
\ee
where
\be
{\cal R}_n\left(\mu(x)\right) := {\cal R}_n\left(\mu(x); a,b,c,d; q\right)
\ee
and
\be
h_n = \dfrac{(q^{c-a-b},q^{d-a},q^{c+d+2},q^{-b};q)_{\infty}}{(q^{{-}a{-}b{-}1},q^{c{+}d{+}1{-}a},q^{c{+}1{-}b},q^{d{+}1};q)_{\infty}}  \dfrac{1-q^{a+b+c+d+2}}{1-q^{a+b+2n+1}} \dfrac{(q,q^{a+b-c+1},q^{a-d+1},q^{b+1};q)_n}{(q^{a+1},q^{a+b+1},q^{b+d+1},q^{c+1};q)_n} \nn
\ee
Therefore, they satisfy a three-term recurrence relation
 \be\label{3}
Y^{(1)}_n{\cal R}_{n+1} + Y^{(2)}_n{\cal R}_{n} + Y^{(3)}_n{\cal R}_{n-1} = x{\cal R}_{n},
\ee
where $Y^{(1,2,3)}_n$ are some coefficients depending on the parameters $a$, $b$, $c$, $d$.
With the help of explicit formulas for these coefficients
(see  \cite[formula (14.2.3)]{koek})
and using relation (\ref{qrac}), one can find
\be
\label{3rel}
0 &=&
a_1 \left\{
\begin{array}{ccc}
r&\bar r&i \\
r&\bar r&j-1
\end{array}
\right\} + a_2 \left\{
\begin{array}{ccc}
r&\bar r&i \\
r&\bar r&j
\end{array}
\right\} + a_3 \left\{
\begin{array}{ccc}
r&\bar r&i \\
r&\bar r&j+1
\end{array}
\right\}
\ee

\be
\label{3rel}
0 &=&
b_1 \left\{
\begin{array}{ccc}
r& r&i \\
\bar r&\bar r&j-1
\end{array}
\right\} - b_2 \left\{
\begin{array}{ccc}
r& r&i \\
\bar r&\bar r&j
\end{array}
\right\} + b_3 \left\{
\begin{array}{ccc}
r& r&i \\
\bar r&\bar r&j+1
\end{array}
\right\}
\ee

with
\be
a_1 &=& [j]^2\,[j-r-1]\,[r+j+N-1]\,[N+2j] \nn\\
a_3 &=& [j-r]\,[j+N-1]^2\,[N+r+j]\,[N+2j-2] \nn\\
a_1+a_2+a_3 &=& -[i]\,[i+N-1]\,[N+2j-2]\,[N+2j-1]\,[N+2j],\\ \nn \\
b_1 &=& [j]^2\,[j+r+1]\,[r-j+N-1]\,[2j+2] \nn\\
b_3 &=& [r-j]\,[j+1]^2\,[N+r+j]\,[2j] \nn\\
b_1+b_2+b_3 &=& [i]\,[i+N-1]\,[2j]\,[2j+1]\,[2j+2].
\ee

\section{Relation to pentagon identities}

In the case of Racah polynomials, the 3-term relations
possess an additional interpretation: they are nothing but the pentagon
(Biedenharn-Elliot) identity \cite{penta,Rbook},
which reflects associativity of the Tanaka-Krein algebra of representations \cite{Rpenta}.
There are five possibilities to decompose into irreducible representations the tensor product $R_1{\otimes}R_2{\otimes}R_3{\otimes}R_4$ of four irreducible representations of the algebra $U_q(\mathfrak{su}_N)$:
\be
\xymatrix{
&[(R_1{\otimes}R_2){\otimes}R_3]{\otimes}R_4 \ar@{<->}[dl] \ar@{<->}[dr]&\\
(R_1{\otimes}R_2){\otimes}(R_3{\otimes}R_4)\ar@{<->}[dd]&&[R_1{\otimes}(R_2{\otimes}R_3)]{\otimes}R_4\ar@{<->}[dd]\\
&&\\
R_1{\otimes}[R_2{\otimes}(R_3{\otimes}R_4)] \ar@{<->}[rr] && R_1{\otimes}[(R_2{\otimes}R_3){\otimes}R_4]\\
}
\ee
One can go over from any decomposition to any other one by a chain clockwise and counterclockwise using the Racah coefficients on each step. Since the final decomposition is the same for the both chains, the matrices of resulting transformations are the same for the both cases. In terms of the 6-j symbols, this equality takes the form
\be
\sum_{R_{34}}\epsilon \cdot\dim_qR_{34}\cdot\left\{
\begin{matrix}R_{12} & R_3 & R_{123} \\ R_4 & R_5 & R_{34} \end{matrix} \right\}
\ \left\{ \begin{matrix} R_1 & R_2 & R_{12} \\ R_{34} & R_5 & R_{234} \end{matrix} \right\} \
\left\{ \begin{matrix} R_{234} & R_2 & R_{34} \cr R_3 & R_4 & R_{23} \end{matrix} \right\}
 \ = \
\left\{ \begin{matrix} R_1 & R_{23} & R_{123} \cr R_4 & R_5 & R_{234} \end{matrix} \right\} \
\left\{ \begin{matrix}R_1 & R_2 & R_{12} \\ R_3 & R_{123} & R_{23} \end{matrix} \right\},
\label{penr} \nn
\ee
where $\epsilon = (-)^{R_{12}+R_{23}+R_{34}+R_{123}+R_{234}-\sum R_i}$, $R_{12} \in R_1\otimes R_2$, $R_{23} \in R_2\otimes R_3$, $R_{34} \in R_3\otimes R_4$, $R_{123} \in R_1\otimes R_2\otimes R_3$, $R_{234} \in R_2\otimes R_3\otimes R_4$. Putting $R_1=R_3=R_{234}=R$, $R_2=R_4=R_{123}=\bar R$, $R_{23}=R_i$, $R_5=R_j$, one obtains
\be
\sum_{R_{34}}\epsilon \cdot\dim_qR_{34}\cdot
\left\{\begin{matrix}R_{12} & R & \bar R \\ \bar R & R_j & R_{34} \end{matrix} \right\}
\ \left\{ \begin{matrix} R & \bar R & R_{12} \\ R_{34} & R_j & R \end{matrix} \right\} \
\left\{ \begin{matrix} R & \bar R & R_{34} \cr R & \bar R & R_i \end{matrix} \right\}
 \ = \
\left\{ \begin{matrix} R & R_i & \bar R \cr \bar R & R_j & R \end{matrix} \right\} \
\left\{ \begin{matrix} R & \bar R & R_{12} \\ R & \bar R & R_i \end{matrix} \right\}.
\label{penr2}
\ee
Hence, $R_{34} \in R\otimes \bar R$ and $R_{34} \in R_j\otimes R_{12}$. Now we can obtain the three-term recurrence relation (\ref{3}) from this formula. To this end, we put $R=[r]$. The Racah matrices we are interested in are $\left\{\begin{matrix}r & \bar r & i \\ r & \bar r & j \end{matrix} \right\}$, where $i$ and $j$ encounter the representations $\mathfrak{R}_n=[2n,n^{N-2}]$.
Let us put also $R_{12}$ to be the adjoint representation of $\mathfrak{su}_N$: $R_{12}=R_{adj} = [2, 1^{N-2}]$. Then, the condition that $R_{34}$ simultaneously belongs to $[r]\otimes [\bar r]$ and to $\mathfrak{R}_j\otimes R_{adj}$ leaves only three term in the sum (\ref{penr2}):
\be
\mathfrak{R}_j\otimes R_{adj} = \mathfrak{R}_{j+1} \oplus \mathfrak{R}_j \oplus \mathfrak{R}_{j-1} \Big |_{[r]\otimes [\bar r]},
\ee
Using the tetrahedral symmetries \cite{GJ}, one obtains
\be
\left\{ \begin{matrix} R & i & \bar R \cr \bar R & j & R \end{matrix} \right\} = \left\{ \begin{matrix} R & \bar R & i \cr R & \bar R & j \end{matrix} \right\}, \
\ \ \ \ \ \ \ \ \ \ \ \ \left\{ \begin{matrix} R & \bar R & R_{34} \cr R & \bar R & i \end{matrix} \right\} = \left\{ \begin{matrix} R & \bar R & i \cr R & \bar R & R_{34} \end{matrix} \right\}.
\ee
Then, the sum (\ref{penr2}) becomes linear in $\mathfrak{R}_*$ and takes the same form as relation (\ref{3}) with some coefficients. In order to find the coefficients, one needs to evaluate the Racah matrices containing the adjoint representation: $\left\{\begin{matrix}R_{12} & R & \bar R \\ \bar R & j & R_{34} \end{matrix} \right\}$, $\left\{ \begin{matrix} R & \bar R & R_{12} \\ R_{34} & j & R \end{matrix} \right\}$, $\left\{ \begin{matrix} R & \bar R & R_{12} \\ R & \bar R & i \end{matrix} \right\}$ as functions of $i$ and $j$.
It is rather tedious to do analytically for arbitrary $N$, but it is
straightforward to extract the coefficients from explicit formulas (\ref{f1})-(\ref{f3}).

\section{q-Racah as $BC$ Koornwinder-Macdonald polynomials}

The Askey-Wilson (q-Racah) polynomials can be described as simplest (one-variable, or one-row) symmetric polynomials for the systems of roots of the $BC_n$ type\footnote{We are grateful to H.Awata for attracting our attention to this essential fact.}. This has been first realized by T.Koornwinder \cite{Koor} (hence, the name Koornwinder-Macdonald polynomials). These symmetric polynomials are constructed from the monomials
\be
m_\lambda=\sum_{\mu\in G(\lambda)} z_1^{\mu_1}z_2^{\mu_2}\ldots z_n^{\mu_s}
\ee
where the sum goes over the orbit $G(\lambda)$ of the partition $\lambda=\{\lambda_1\ge\lambda_2\ge\ldots\ge 0\}$ under the action of the group $G=S_s\times \mathbb{Z}_2$ which permutes $\lambda_i$ and changes their signs. This is nothing but the $BC_s$-type Weyl group. Now defining the second order difference operator \cite{Koor}
\be\label{difop}
\hat{\mathfrak{D}}=\sum_j\left(P(z;z_j)(\hat T_j-1)+P(z;z_j^{-1})(T_j^{-1}-1)\right)
\ee
where
\be
P(z;z_j)={\prod_{a=0}^3 (1-t_az_j)\over (1-z_j^2)(1-qz_j^2)}\prod_{k\ne j}{(1-tz_jz_k)(1-tz_jz_k^{-1})\over (1-z_jz_k)(1-z_jz_k^{-1})},\ \ \ \ \ \ \
\hat T_j f(z_1,z_2,\ldots)=f(z_1,z_2,\ldots,qz_j,\ldots)
\ee
one can construct the set of $BC$-Macdonald (or Koornwinder-Macdonald) polynomials
\be
p_\lambda=\left(\prod_{\mu\le\lambda}{\hat{\mathfrak{D}}-E_\mu\over E_\lambda-E_\mu}\right)m_\lambda
\ee
where
\be
E_\lambda=\sum_{j=1}^s q^{-1}t_0t_1t_2t_3t^{2n-j-1}(q^{\lambda_j}-1)+t^{j-1}(q^{-\lambda_j}-1)
\ee
and $\mu\le\lambda$ is understood as $\sum_{j=1}^k(\lambda_j-\mu_j)\ge 0$ for all $k=1,\ldots,s$.
These polynomials are the eigenfunctions of the difference operator $\hat{\mathfrak{D}}$ (\ref{difop}) with the eigenvalues $E_\lambda$:
\be
\hat{\mathfrak{D}}p_\lambda=E_\lambda p_\lambda
\ee
Considering the case of one variable $z$, i.e. $s=1$, which is equivalent to one-line Young diagram $\lambda$ corresponding to symmetric representations, one arrives at the Askey-Wilson polynomials (\ref{AV}) with $z=e^{i\theta}$ and $a=t_0$, $b=t_1$, $c=t_2$, $d=t_3$. Note that, in this case, the $t$-dependence automatically disappears from the polynomial.

\bigskip

Note that there is also another multivariable generalization of the q-Racah polynomials \cite{Trat,GR2,Il},
\be
{\cal R}_{\bf n}\left(\mu({\bf x}),\Big|\,{\bf a},b,c,\Big|\,q\right)= \nn \\
= \prod_{k=1}^s{\cal R}_{n_k}
\left(\mu(x_k-N_{k-1})\,\Big|\,
b+A_k+2N_{k-1}-a_1,\ a_{k+1}-1,\ A_k+x_{k+1}+N_{k-1},\ x_{k+1}-N_{k-1}\,\Big|\,q\right)
\ee
where
\be
A_0=0,\ \ \ \ A_k=\sum_{j=1}^kA_j,\ \ \ \ x_{s+1}=-c-1,\ \ \ \ N_0=0,\ \ \ \ N_k=\sum_{j=1}^kn_j
\ee
and ${\bf n}$, ${\bf x}$ denote the sets of $s$ integers $\{n_i\}$ and of $s$ variables $\{x_i\}$ correspondingly, while ${\bf a}=\{a_1,a_2,\ldots,a_{s+1}\}$. Here also $N_s\le x_{s+1}$.
These polynomials are still orthogonal.

\section{Conclusion}

In this paper, we expressed the analytic formulas for Racah matrices in all
symmetric (and, hence, antisymmetric) representations through terminating balanced hypergeometric series,
that is, through the orthogonal  Askey-Wilson (Racah) $q$-polynomials.
Our main result is extension of the Racah-matrix interpretation
from the 1-parametric sub-family of such polynomials
to two 2-parametric ones, by introducing the second parameter $N$ (from $\mathfrak{su}_N$)
in addition to $r$, which describes the symmetric representation.
At the same time, the entire variety of the Askey-Wilson polynomials is 4-parametric,
and three extra parameters in the case of  $\mathfrak{su}_2$ are associated with
four different symmetric
representations characterized by four different parameters $r_1,r_2,r_3,r_4$:
\be
\left\{
\begin{array}{ccc}
r_1& r_3&i \\
r_2&r_4&j
\end{array}
\right\} \sim{\cal R}_{\frac{1}{2}(r_1+r_3)-j}{\footnotesize
\left(q^{2i{-}r_1{-}r_3}{+}q^{{-}r_1{-}r_3{-}2(i{+}1)}\,\Big|\,{-}r_3{-}1,{-}r_2{-}1,{-}\frac{1}{2}(r_1{+}r_2{+}r_3{+}r_4){-}2,\frac{1}{2}(r_2{-}r_1{+}r_4{-}r_3)
\,\Big|\,q^2\right)}
\nn
\ee
% (this is one of the many symmetry related forms of the formula, which is not fully correlated with the choice in the main text).
Since this case uses all available parameters the orthogonal polynomials described by the balanced terminating hypergeometric series $_4\phi_3$, one could expect that extending this formula to the $\mathfrak{su}_N$ case leads to higher hypergeometric series, since $N$ will be an {\it additional} parameter. However, it turns out that, in the case of two coinciding symmetric representations and two their $\mathfrak{su}_N$-conjugated, the same hypergeometric series is still sufficient.

The next problem for Racah calculus is to go beyond the symmetric representation,
in particular to find analytic expressions for already known Racah matrices
$S$ and $\bar S$ in various {\it two}-line representations, especially, in the
rectangular ones, where there are no multiplicities and no associated ambiguities
with the choice of bases in arborescent calculus \cite{arbor}.
Direct attempts to guess such formulas as interpolating between the known
matrix elements ${\bar S}_{ij}$ for particular $i$ and $j$ are somewhat tedious,
especially because the number of summations is unknown.
At the same time, in hypergeometric and Macdonald calculi, there are natural ways
for generalizations to higher representations, and this can significantly
simplify the problem.

Also, the hypergeometric functions possess integral representations,
which can be interpreted \cite{MV} as correlators of conformal blocks within the Dotsenko-Fateev
formalism \cite{DF,AGTmamo,AGTmamo5d}, which is conceptually interesting.
We remind that the Racah matrices naturally describe modular transformations
of the conformal blocks, while the fact that matrices of the transformations of some objects
can be also considered as the same objects themselves is intriguing and promising.
A good example of this phenomenon is given by the celebrated $RTT$ relations,
where the $R$-matrix itself coincides with the group element $T$ in a proper representation of the algebra of functions.
Another direction suggested by relation to the conformal blocks, is an additional $t$-deformation,
from matrix to network models {\it a la} \cite{Nagoya}, i.e. from quantum groups
to DIM algebras.

Anyhow, even without these additional bonuses,
expression through Askey-Wilson polynomials obtained in the present letter
provides a very compact
(most economic) and elegant description of the Racah matrices, and has a value of its own.

\section*{Acknowledgements}
This work was funded by the Russian Science Foundation (Grant No.16-11-10291).


\begin{thebibliography}{99}

\bibitem{RW}  G. Racah, %Theory of Complex Spectra. II.
Phys.Rev. {\bf 62} (1942) 438-462\\
E.P. Wigner, Manuscript in 1940, appeared in: {\sl Quantum Theory of Angular Momentum},
pp. 87–133. (Academic Press, New York 1965);
{\sl Group Theory and Its Application to the Quantum Mechanics of Atomic
Spectra} (Academic Press, New York 1959)

\bibitem{LL3}  L.D. Landau and E.M. Lifshitz, {\sl Quantum Mechanics: Non-Relativistic Theory},
3rd ed., Pergamon Press, (1977)

\bibitem{Rbook} J. Scott Carter, D.E. Flath, M. Saito, {\sl The Classical and Quantum 6j-symbols}, Princeton University Press, 1995

\bibitem{Racah} S. Nawata, P. Ramadevi and Zodinmawia, Lett.Math.Phys. {\bf 103} (2013) 1389-1398, arXiv:1302.5143

\bibitem{GJ} J. Gu and H. Jockers, arXiv:1407.5643

\bibitem{MMSpret} A. Mironov, A. Morozov and A. Sleptsov, JHEP {\bf 07} (2015) 069,  arXiv:1412.8432

\bibitem{MMMS} A. Mironov, A. Morozov, An. Morozov and A. Sleptsov, JETP Lett. {\bf 104} (2016) 56-61 (Pisma Zh.Eksp.Teor.Fiz. {\bf 104} (2016) 52-57), arXiv:1605.03098\\
A. Mironov, A. Morozov, An. Morozov and A. Sleptsov, Physics Letters {\bf B760} (2016) 45-58, arXiv:1605.04881

\bibitem{Mor} A. Morozov, JHEP {\bf 1609} (2016) 135, arXiv:1606.06015; arXiv:1612.00422; Phys.Lett. {\bf B766} (2017) 291-300,  arXiv:1701.00359

\bibitem{knotebook} http://www.knotebook.org

\bibitem{Kaul} P. Ramadevi, T.R. Govindarajan and R.K. Kaul,
Mod.Phys.Lett. {\bf A9} (1994) 3205-3218, hep-th/9401095

\bibitem{NRZ} S. Nawata, P. Ramadevi and Zodinmawia,
J.Knot Theory and Its Ramifications {\bf 22} (2013) 13, arXiv:1302.5144\\
Zodinmawia's PhD thesis, 2014

\bibitem{GMMMS} D. Galakhov, D. Melnikov, A. Mironov, A. Morozov and A. Sleptsov, Phys.Lett. {\bf B743} (2015) 71-74, arXiv:1412.2616\\
D. Galakhov, D. Melnikov, A. Mironov and A. Morozov, Nucl.Phys. {\bf B899} (2015) 194-228, arXiv:1502.02621

\bibitem{arbor}  A.Mironov, A.Morozov, An.Morozov, P.Ramadevi and Vivek Kumar Singh,
JHEP {\bf 1507} (2015) 109,  arXiv:1504.00371\\
A.Mironov and A.Morozov, Phys.Lett. B755 (2016) 47-57 arXiv:1511.09077 \\
A. Mironov, A. Morozov, An. Morozov, P. Ramadevi, Vivek Kumar Singh and A. Sleptsov,  J.Phys. A: Math.Theor. {\bf 50} (2017) 085201, arXiv:1601.04199

\bibitem{IMMMfe}
H. Itoyama, A. Mironov, A. Morozov and An. Morozov, JHEP {\bf 7} (2012), 131, arXiv:1203.5978

\bibitem{diffexpan} N.M. Dunfield, S. Gukov and J. Rasmussen, Experimental Math. {\bf 15} (2006) 129-159, math/0505662\\
S.Gukov and M.Stosic,  arXiv:1112.0030\\
E.Gorsky, S.Gukov and M.Stosic, arXiv:1304.3481\\
A. Mironov, A. Morozov and An. Morozov, AIP Conf. Proc. {\bf 1562} (2013) 123, arXiv:1306.3197\\
S. Arthamonov, A. Mironov, A. Morozov and An. Morozov,  JHEP {\bf 04} (2014) 156,  arXiv:1309.7984\\
S. Gukov, S. Nawata, I. Saberi, M. Stosic and P. Sulkowski, arXiv:1512.07883\\
A.Morozov, Nucl.Phys. B911 (2016) 582-605,  arXiv:1605.09728;\\
Ya.Kononov and A.Morozov,   arXiv:1609.00143;
Mod.Phys.Lett. {\bf A31} (2016) 1650223,  arXiv:1610.04778

\bibitem{Sulk}  P. Kucharski, M. Reineke, M. Stosic and P. Sulkowski,  arXiv:1707.02991;  arXiv:1707.04017

\bibitem{AW} J. Wilson, {\sl Hypergeometric series recurrence relations and some new orthogonal functions}, Ph.D. thesis (1978), Univ. Wisconsin, Madison\\
R. Askey, J. Wilson,
%{\sl  A set of orthogonal polynomials that generalize the Racah coefficients or 6j-symbols},
SIAM Journal on Mathematical Analysis {\bf 10} (1979) 1008-1016

\bibitem{KK} I.I. Kachurik and A.U. Klimyk,
%{\it On Racah coefficients of the quantum algebra $U_q(su_2)$},
J.Phys. A: Math.Gen. {\bf 23} (1990) 2717

\bibitem{Klimyk-Vilenkin} N. Ja. Vilenkin and A. U. Klimyk, {\sl Representation of Lie Groups and Special Functions. Volume 1: Simplest Lie Groups, Special Functions and Integral Transforms}, Springer Science+Business Media Dordrecht, 1991

\bibitem{Kirilov-Reshetikhin} A. Kirillov and N. Reshetikhin,
%{\it Representations of the algebra $U_q(sl(2))$, q-orthogonal polynomials
%and invariants of Links.}
In:
%Kohno, T. (ed.)
{\sl New Developments in the Theory of Knots},
World Scientific, Singapore (1989)

\bibitem{penta} L.C. Biedenharn,  J.Math.Phys. (MIT) {\bf 31} (1953) 287;
J.P. Elliott,  Proc.R.Soc. {\bf A218} (1953) 370

\bibitem{Mac}  I.G. Macdonald, {\sl Orthogonal polynomials associated with root systems}, unpublished manu-
script, 1988 (S\'eminaire Lotharingien Combin. {\bf 45} (2000), Article B45a, 40 pp, math/0011046)

\bibitem{Koor} T. H. Koornwinder, {\sl Askey-Wilson polynomials for root systems of type BC}, in: {\sl Hyperge-
ometric functions on domains of positivity, Jack polynomials, and applications} (D. St. P.
Richards, ed.), Contemp.Math. {\bf 138}, Amer.Math.Soc., Providence, R.I., 1992, pp.
189–204

\bibitem{GR} G. Gasper and M. Rahman, {\sl Basic hypergeometric series},
Cambridge University Press, 1990

\bibitem{koek} R. Koekoek, P.A. Lesky and R.F. Swarttouw,
{\sl  Hypergeometric orthogonal polynomials and their q-analogues},
Springer Science $\&$ Business Media, 2010

\bibitem{Rpenta} P. Etingof, S. Gelaki, D. Nikshych, V. Ostrik, {\sl Tensor Categories} (2009), http://www-math.mit.edu/~etingof/tenscat1.pdf

\bibitem{Trat} M.V. Tratnik,
%{\sl Some multivariable orthogonal polynomials of the Askey tableau-continuous families},
J.Math.Phys. {\bf 32} (1991) 2065–2073

\bibitem{GR2} G. Gasper and M. Rahman, %Some systems of multivariable orthogonal q-Racah polynomials,
Ramanujan J. {\bf 13} (2007) 389–405, arXiv:math/0410250

\bibitem{Il} P. Iliev,  Trans.Amer.Math.Soc. {\bf 363} (2011) 1577-1598, arXiv:0801.4939

\bibitem{MV}
A. Morozov and L. Vinet, Mod.Phys.Lett. {\bf A8} (1993) 2891-2902, arXiv:hep-th/9309026 \\
A. Mironov, A. Morozov and L. Vinet, Theor.Math.Phys. {\bf 100} (1995) 890-899, arXiv:hep-th/9312213

\bibitem{DF} Vl. Dotsenko and V. Fateev, Nucl.Phys. {\bf B240} (1984) 312-348

\bibitem{AGTmamo}
R. Dijkgraaf and C. Vafa, arXiv:0909.2453;\\
H. Itoyama, K. Maruyoshi and T. Oota,
%\emph{Notes on the Quiver Matrix Model and 2d-4d Conformal Connection},
Prog.Theor.Phys. {\bf 123} (2010) 957-987, arXiv:0911.4244\\
T. Eguchi and K. Maruyoshi,
%\emph{Penner Type Matrix Model and Seiberg-Witten Theory},
arXiv:0911.4797;
%{\it Seiberg-Witten theory, matrix model and AGT relation},
arXiv:1006.0828\\
R. Schiappa and N. Wyllard,
%\emph{An $A_r$ threesome: Matrix models, $2d$ CFTs and $4d$ N=2 gauge theories},
arXiv:0911.5337\\
A. Mironov, A. Morozov and Sh. Shakirov,
%\emph{Matrix Model Conjecture for Exact BS Periods and Nekrasov Functions},
JHEP {\bf 02} (2010) 030, arXiv:0911.5721;
%\emph{Conformal blocks as Dotsenko-Fateev Integral Discriminants},
Int.J.Mod.Phys. {\bf A25} (2010) 3173-3207, arXiv:1001.0563\\
H. Itoyama and T. Oota, Nucl. Phys. {\bf B838} (2010) 298-330, arXiv:1003.2929\\
A. Mironov, A. Morozov and An. Morozov, Nucl.Phys. {\bf B843} (2011) 534-557, arXiv:1003.5752

\bibitem{AGTmamo5d} H.~Awata and H.~Kanno,
  %``Quiver Matrix Model and Topological Partition Function in Six Dimensions,''
  JHEP {\bf 0907} (2009) 076
%  doi:10.1088/1126-6708/2009/07/076
arXiv:0905.0184\\
H.~Awata and Y.~Yamada,
  %``Five-dimensional AGT Conjecture and the Deformed Virasoro Algebra,''
  JHEP {\bf 1001} (2010) 125,
%  doi:10.1007/JHEP01(2010)125
arXiv:0910.4431;
  %``Five-dimensional AGT Relation and the Deformed beta-ensemble,''
  Prog.\ Theor.\ Phys.\  {\bf 124} (2010) 227,
%  doi:10.1143/PTP.124.227
arXiv:1004.5122\\
S. Yanagida, J.Math.Phys. {\bf 51} (2010) 123506 arXiv:1005.0216\\
A. Mironov, A. Morozov, S. Shakirov and A. Smirnov, Nucl. Phys. {\bf B855} (2012) 128, arXiv:1105.0948\\
  F. Nieri, S. Pasquetti, F. Passerini and A. Torrielli, JHEP {\bf 12} (2014) 040, arXiv:1312.1294\\
M.-C. Tan, JHEP {\bf 12} (2013) 031, arXiv:1309.4775; arXiv:1607.08330\\
H. Itoyama, T.Oota and R. Yoshioka, J.Phys. A: Math.Theor. {\bf 49} (2016) 345201, arXiv:1602.01209\\
% \bibitem{6dAGT}
  A. Nedelin and M. Zabzine, arXiv:1511.03471\\
    Y.~Ohkubo, H.~Awata, H.~Fujino,
  %``Crystallization of deformed Virasoro algebra, Ding-Iohara-Miki algebra and 5D AGT correspondence,''
  arXiv:1512.08016

\bibitem{Painleve} A. Mironov and A. Morozov, Phys.Lett. {\bf B773} (2017) 34-46, arXiv:1707.02443; arXiv:1708.07479

\bibitem{Nagoya} A.~Mironov, A.~Morozov and Y.~Zenkevich, Phys.Lett. {\bf B762} (2016) 196-208, arXiv:1603.05467\\
H. Awata, H. Kanno, T. Matsumoto, A. Mironov, A. Morozov, An. Morozov, Y. Ohkubo and Y. Zenkevich, JHEP {\bf 07} (2016) 103, arXiv:1604.08366



\end{thebibliography}
\end{document}